\documentstyle[11pt]{article}
\input{epsf}
\def\dofig#1#2{\epsfysize=#1 \centerline{\epsfbox{#2}}}

\begin{document}

\title{Delayed Dynamics toward Applications}
\author{Toru Ohira\\
Sony Computer Science Laboratory,\\
3-14-13, Higashigotanda, Shinagawa, Tokyo, 141 Japan\\
ohira@csl.sony.co.jp\\
}
\date{\  }
\maketitle

\begin{abstract}
We propose two dynamical models with delay taking advantage of
their complex dynamics for information processing tasks. The first model incorporates coupled delayed dynamics of multiple bits, which is shown to have desirable properties as an encryption scheme. The second model is a single binary element with delayed stochastic transition, which
presents a resonance behavior between noise and delay.
\footnote{To appear in the proceedings of 1998 International Symposium
 on Nonlinear
Theory and its Applications (Nolta'98)
 September, 1998, 
Crans-Montana, Switzerland }
\end{abstract}

{\bf \section{INTRODUCTION}}
Complex behaviors due to time delays
are found in many natural and artificial systems. 
Some examples are delays in bio-physiological
controls \cite{mackey,longtin}, and signal transmission delays 
in large--scale networked or distributed information
systems (See e.g. \cite{jalote,jha}). 
Research on systems or models with delay has
also been carried out in the fields of 
mathematics \cite{cooke,kuchler}, artificial neural networks \cite{marcus,hertz},
and in physics \cite{derstine,pyragas,ohira}.
This series of research has revealed that time delay can
introduce surprisingly complex behaviors to otherwise
simple systems, because of which delay
has been considered an obstacle from the
point of view of information processing.

In this paper, however, we actually take advantage of this complexity
with delayed dynamics and propose two models. The first model incorporate
delayed dynamics into a new model of encryption. The second model
shows a resonance behavior between noise and delay. Through these two models, we explore the possibility of applications for delayed dynamics.

\section{THE ENCRYPTION MODEL}
The encryption process is identified with 
a coupling dynamics with various time delays between different
bits in the original data. 
We show that the model produces a  complex behavior with
the characteristics needed for an encryption scheme.

Let us now describe the encryption model in more detail. 
${\bf{S}}(0)$ is the original data of $N$ binary bits, whose
$i$th element $s_i(0)$ takes values $+1$ or $-1$.
The delayed dynamics for the encryption can be specified
by a key which consists of the following three parts:  
(1) a permutation ${\bf P}$ generated from $(1,2,3,..,N)$, 
(2) a delay parameter vector \mbox{\boldmath $\tau$} which consists of $N$ positive
integers, and 
(3) number of iterations of the dynamics $T$.
Given the key $K = ({\bf P}$, \mbox{\boldmath $\tau$}, $T)$, the dynamics
is defined as
\begin{equation}
s_i(t) = (-1)\times s_{p_i}(t-\tau_i), 
\end{equation}
where $p_i$ and $\tau_i$ are $i$th element of ${\bf P}$ and \mbox{\boldmath $\tau$}, 
respectively. (If $t-\tau_i < 0$, we set $t - \tau_i = 0$.)
In Figure 1, this dynamics is shown schematically.
The state of the $i$th element of ${\bf{S}}(t)$ is given
by flipping the state of the $p_i$th element of ${\bf{S}}(t-\tau_i)$.
Thus this dynamics causes interaction between $N$ bits of 
the data in both space and time. The encoded state 
${\bf{S}}(T)$ is obtained by applying this operation
of equation (1) iteratively $T$ times starting from
${\bf{S}}(0)$.
\begin{figure}[h]
\dofig{3.5cm}{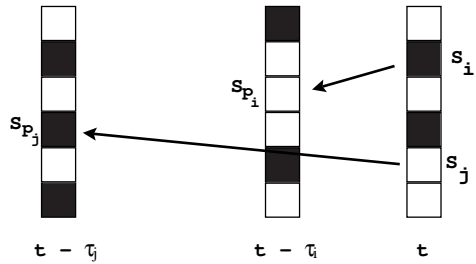}
\caption{
Schematic view of the model dynamics. 
The state of the $i$th element of ${\bf{S}}(t)$ is given
by flipping the state of the $p_i$th element of ${\bf{S}}(t-\tau_i)$.
} 
\label{fig1} 
\end{figure}

We investigate numerically the nature of the delayed dynamics 
from the perspective
of measuring the strength as an encryption scheme.
First, we examine how the state ${\bf{S}}(t)$ evolves
with time. In Figure 2 (A), we have shown an
example of encoded states with different $T$ using the
same ${\bf P}$ and \mbox{\boldmath $\tau$} for a case of $N=81$.   
To be more quantitative,
we compute the following quantity as a measure of difference
between two encoded states at different times $t$ and $t_f$:
\begin{equation}
Y(t)={1 \over N} \, \sum_{i=0}^NS_i(t)S_i\left(t_f\right)
\end{equation}

A typical example is shown in Figure 2(B). We note 
that the dynamics of our model has a property of
occasionally very similar, but not exactly the same, states appearing
indicated by sharp peaks in the figure.
Except around these particular points, however,  we generally
obtain rather uncorrelated encoded states (i.e., $Y \approx 0$) with
different iteration times. This is a desirable property of the model as
an encryption scheme: the same state can be encoded into
uncorrelated states by changing $T$.
\begin{figure}[h]
\dofig{9.3cm}{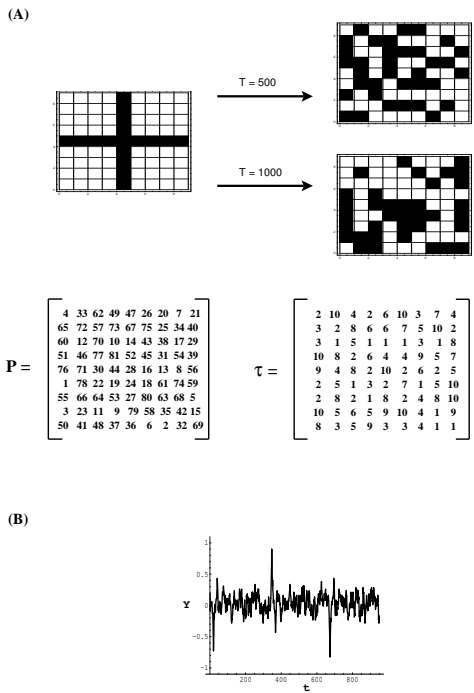}
\caption{
(A) Examples of encoding with the model dynamics from an initial state to 
$T=500$ and $T=1000$ with ${\bf P}$ and  $\tau$.
(B) Example of the correlation $Y$ between encoded states at different time steps evaluated by
equation (2).  Cases with $t_f = 1000$ are plotted
with the initial state and ${\bf P}$ and $\tau$ the same as in Figure 2 (A).
} 
\label{betavary} 
\end{figure}

Next, we investigated the effect of a minor change of ${\bf P}$ and \mbox{\boldmath $\tau$}
on the model dynamics. Starting with the same initial condition,
we evaluate how two states ${\bf{S}}(t)$ and ${\bf{S'}}(t)$ are encoded with slightly different 
${\bf P}$ and ${\bf P'}$, respectively, by computing
\begin{equation}
X(t)={1 \over N} \, \sum_{i=0}^NS'_i(t)S_i\left(t\right)
\end{equation}
A representative result is shown in Figure 3(A).  The same evaluation with
\mbox{\boldmath $\tau$} and \mbox{\boldmath $\tau' $} is shown in Figure 3(B).
These graphs indicate that if we take sufficiently large $T$, the
same state can evolve into rather uncorrelated states
with only a slight change of ${\bf P}$ and \mbox{\boldmath $\tau$}. 
This again is a favorable property in the light of encryption.
It makes iterative and gradual guessing of ${\bf P}$ and \mbox{\boldmath $\tau$}
in terms of their parts and elements very difficult: a nearly correct guess
of the values of ${\bf P}$ and \mbox{\boldmath $\tau$} does not help in decoding.
\begin{figure}[h]
\dofig{5.5cm}{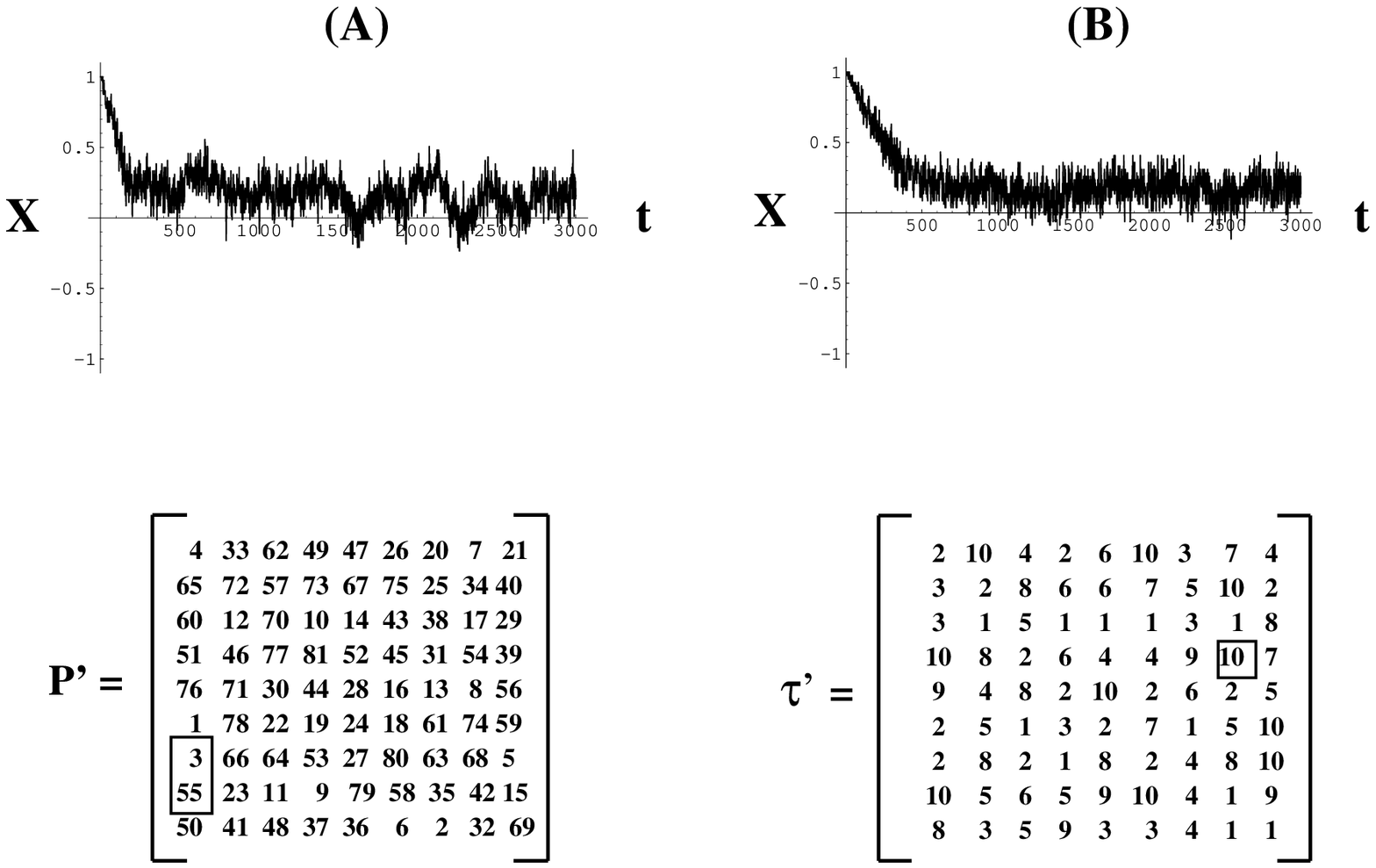}
\caption{
Example of the correlation $X$ evaluated by
equation (3) between two states encoded by slightly different
(A) ${\bf P}$ and ${\bf P'}$, and (B) $\tau$ and $\tau' $.  The difference of ${\bf P'}$ from ${\bf P}$, 
and $\tau' $ from $\tau$
are indicated by boxes. 
The initial state and ${\bf P}$ and $\tau$ are the same as in Figure 2.
} 
\label{fig3} 
\end{figure}

With these properties of the model, an exhaustive search appears to be
the only method for guessing the key.  Even if  one knows $N$ and 
$\tau_{max}$, the largest element in \mbox{\boldmath $\tau$},  one is still required to
search for the correct key from among $(N!)(\tau_{max})^N$ combinations and to guess $T$. 
The commonly used DES (Data Encryption Standard) employs $2^{56}$ bit
keys \cite{des}. We can obtain a similar order of difficulty 
with rather small values of $N$ and $\tau_{max}$; 
for instance,  $N \approx 11$ and $\tau_{max} \approx10$ \cite{note1}. 

There are different methods possible for using this model for a secure communication
between two persons who share the key.  One example is that
the sender sends a series of encoded data in sequence for the interval
between $T$ and $T+\tau_{max}$ (or longer). 
The receiver can recover 
the original data from this set of encoded data by applying a reverse
dynamics with the key.   In a situation where the data sent is a choice out
of multiple data sets known to the receiver,  the receiver can
run the encryption dynamics to the entire sets with the key for
case matching.

\section{RESONANCE BETWEEN NOISE AND DELAY}
The second model is a stochastic binary element with delayed
transition which is schematically 
shown in Figure 4. The state of the element $X(t)$ at time
step $t$ can take either $-1$ or $1$ and the transition probabilities
are indicated by the arrows. We further include delay in the model,
which makes the formal definition of the model as follows:
\begin{eqnarray}
P(1, t+1) &=& p\ \ \quad X(t-\tau) = -1,\nonumber\\
               &=& 1-q \quad X(t-\tau) = 1,\\
P(-1, t+1) &=& q \ \ \quad X(t-\tau) = 1,\nonumber\\
                &=& 1-p \quad X(t-\tau) = -1,
\end{eqnarray}
where $P(s, t)$ is a probability that $X(t)=s$.  
\begin{figure}[b]
\dofig{4cm}{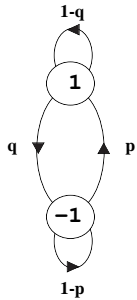}
\caption{
Schematic view of the stochastic model dynamics. 
} 
\label{fig4} 
\end{figure}
Hence, the transition probability
of this model depends on its state at $\tau$ steps earlier. In this sense,
this model is a special case of "delayed random walks"\cite{ohira,ohira1} except it can 
take only two states.

Let us now discuss our motivation for considering such a model.
It has been shown that a simple stochastic binary element can 
show a resonant behavior with a external oscillating signal with an appropriate choice of transition
probability. This phenomena is termed as ``Stochastic Resonanc'' (see e.g. \cite{bulsara}) 
 now widely investigated. In our model here, we have replaced an external oscillatory
signal with delay and its associated oscillatory behavior. Hence, we can expect
some form of resonance can be seen with the model between the oscillatory
dynamics due to noise and that due to delay. 

A preliminary simulation study shows this is indeed the case. 
We have fixed $q=1-q=0.5$ and $\tau=10$, and varied $p$. 
The simulations are done up to 20000 steps.
In Figure 5, where we have plotted examples of residence time histograms  in
the $-1$ state and dynamics of $X(t)$, we can see a resonance
phenomena as a height of the peak at the period equal to the delay. 
If we tune the noise ($p$) appropriately,
we observe the height of the residence time histogram peak reaching
the maximum value. This is shown in Figure 6\cite{note2}.

With this property, this model can be used to stochastically encode 
information. We have built a preliminary application to encode
static pictures using this model, which will be reported elsewhere.

\section{Summary}
We have proposed two models which include delay.
These two models indicate that complex delayed dynamics can be used for
information processing tasks rather than necessarily being obstacle. More thorough investigation
of each model is presented elsewhere, and concrete applications
 of these models are 
currently being devloped.
\clearpage

\begin{figure}[h]
\dofig{16cm}{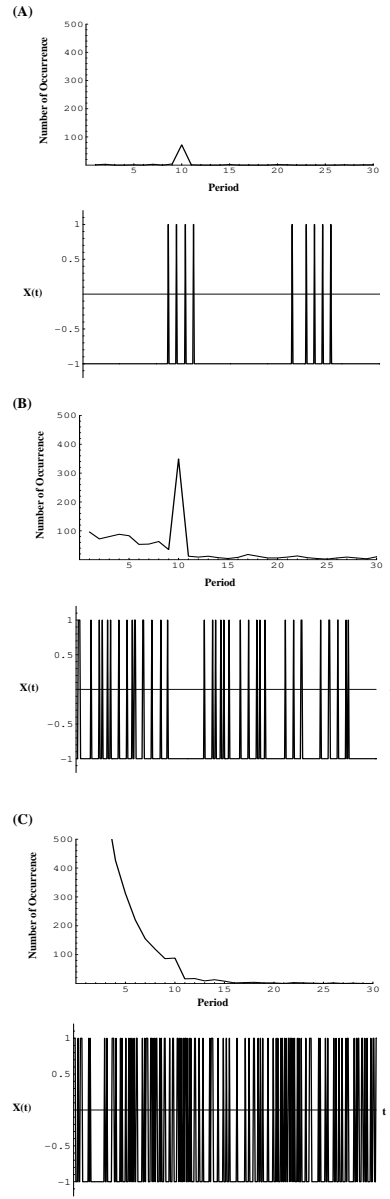}
\caption{
The residence time histogram and dynamics of $X(t)$ as we change $p$.
The values of $p$ are (A) $p=0.005$, (B) $p=0.04$, (C) $p=0.2$.
} 
\label{fig5} 
\end{figure}
\clearpage

\begin{figure}[h]
\dofig{3cm}{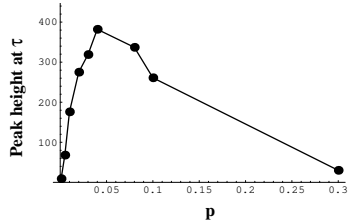}
\caption{
A plot of peak height with varying $p$
} 
\label{fig6} 
\end{figure}


\begin{thebibliography}{99}
\bibitem{note2}
For the case of fixed $p$ and varying $\tau$, a preliminary
investigation shows that a monotone decreasing
histograms appear for $\tau > 0$.
\end{thebibliography}
\end{document}